\documentclass[english,superscriptaddress,reprint,two column]{{revtex4}}
\usepackage[T1]{fontenc}
\usepackage{color}
\usepackage{amsmath}
\usepackage{amssymb}
\usepackage{graphicx}
\usepackage{amsfonts}
\usepackage{transparent}
\usepackage{babel}
\usepackage{bm}
\usepackage{epsfig}
\usepackage{enumerate}
\usepackage{multirow}
\usepackage{verbatim}
\usepackage{float}
\usepackage[breaklinks=true,colorlinks=true,linkcolor=blue,urlcolor=blue,citecolor=blue]{hyperref}
\usepackage{caption}


\renewcommand{\raggedright}{\leftskip=0pt \rightskip=0pt plus 0cm}
\begin{document}

\title{Non-Hermitian skin effect and nonreciprocity induced by dissipative couplings}

\author{Xinyao Huang}
\affiliation{State Key Laboratory of Low-Dimensional Quantum Physics, Department of
Physics, Tsinghua University, Beijing 100084, China}
\affiliation{School of Physics, Beihang University, Beijing 100191, China}

\author{Yaohua Li}
\affiliation{State Key Laboratory of Low-Dimensional Quantum Physics, Department of
Physics, Tsinghua University, Beijing 100084, China}

\author{Guo-Feng Zhang}
\affiliation{School of Physics, Beihang University, Beijing 100191, China}

\author{Yong-Chun Liu}
\email{ycliu@tsinghua.edu.cn}
\affiliation{State Key Laboratory of Low-Dimensional Quantum Physics, Department of
Physics, Tsinghua University, Beijing 100084, China}
\affiliation{Frontier Science Center for Quantum Information, Beijing 100084, China}

\begin{abstract}
    We study the mechanism for realizing non-Hermitian skin effect (NHSE) via dissipative couplings, in which the left-right couplings have equal strengths but the phases do not satisfy the complex conjugation. Previous realizations of NHSE typically require unequal left-right couplings or on-site gain and loss. In this work we find that when combined with the multichannel interference provided by a periodic dissipative-coherent coupling structure, the dissipative couplings can lead to unequal left-right couplings, inducing NHSE. Moreover, we show that the non-Hermiticity induced by dissipative couplings can be fully transformed into nonreciprocity-type non-Hermiticity without bringing extra gain-loss-type non-Hermiticity. Thus, this mechanism enables unidirectional energy transmission without introducing additional insertion loss. Our work opens a new avenue for the study of non-Hermitian topological effects and the design of directional optical networks.

\end{abstract}
    
    \maketitle
    
\textit{Introduction.---}
Non-Hermitian systems bring unprecedented features induced by non-Hermiticity~\cite{Bender2007,Bergholtz2021, Ashida2020,Zhang2022a}. For instance, non-Hermitian systems with parity-time$ (\mathcal{PT}$) symmetry can possess real spectra~\cite{Bender1998, Regensburger2012,Konotop2016,Feng2017,El-Ganainy2018,Ozdemir2019,Xia2021}. The $\mathcal{PT}$ phase transition occurs at the non-Hermitian exceptional point~\cite{Heiss2004}, which describes the coalescence of the eigenstates and the degeneracy of the eigenvalues~\cite{Wiersig2014,Hodaei2017,Chen2017,Ozdemir2019}. As a most peculiar example, non-Hermitian skin effect (NHSE) driven by non-Hermiticity has drawn much attention in recent years~\cite{Yao2018,Kunst2018,Martinez-Alvarez2018,Yokomizo2019,Edvardsson2019,Lee2019,Borgnia2020,Okuma2020,Yang2020,Roccati2021,Wang2022,Ghatak2020,Garbe2023}. It describes that the majority of eigenstates are localized at the boundaries, implying the breakdown of the conventional bulk-boundary correspondence (BBC) in Hermitian systems~\cite{Yao2018a,Song2019,Xiao2020,Helbig2020,Zhang2020}. To date, NHSE has been proposed in various setups such as optical \cite{Xiao2020,Weidemann2020} and acoustic systems~\cite{Zhang2021,Zhang2021a}, cold atoms~\cite{Liang2022}, circuits~\cite{Helbig2020,Zou2021}, and bosonic systems governed by quadratic Hamiltonians~\cite{Xu2021,Wang2022a,Yokomizo2021,Yang2020a,Wang2019}. In general, the realization of NHSE is mainly based on two origins of non-Hermiticities: unequal left-right couplings~\cite{Yao2018,Lee2019} and on-site gain and loss (or unequal on-site losses)~\cite{Lee2016,Li2022}. 
    
As a different origin of non-Hermiticity, dissipative couplings connect the systems indirectly via a non-Hermitian reservoir, indicating the irreversible energy flow in the coupling channel~\cite{Krauter2011,Vasilyev2013,Metelmann2015,Mukherjee2017,Ding2019}. 
Up to now, dissipative couplings have been realized in various setups, such as optical cavities~\cite{Zhang2020a,Arwas2022}, thermal atomic ensembles~\cite{Krauter2011,Peng2016}, micromechanical oscillators~\cite{Fang2017}, and optical fibres~\cite{Bergman2021}. 
In addition to being a different class of coupled networks, the non-Hermitian properties exhibited by dissipatively coupled systems make them promising platforms for studying unconventional topological properties~\cite{Malzard2015,Leefmans2022,Hao2023}. However, the relationship between dissipative couplings and NHSE remains unknown.

Here we show that the presence of dissipative couplings can induce NHSE as well as nonreciprocal energy transmission. Taking a chain of resonance modes with intracell dissipative couplings and intercell coherent couplings as an example, we find that NHSE can be realized by combining dissipative couplings with multichannel interference provided by the chain. The system can be transformed into the non-Hermitian Su-Schrieffer-Heeger (SSH) model with unequal left-right couplings, and enables nonreciprocal energy transmission.
Distinct from the non-Hermitian Hamiltonians of the systems with unequal left-right couplings and systems with on-site gain and loss, we find that the non-Hermitian Hamiltonians including dissipative couplings preserve local anti-$\mathcal{PT}$ symmetry~\cite{Peng2016,Yang2017,Arwas2022}. 
In the aspect of nonreciprocity generation, the mechanism proposed here is also different from the schemes based on tuning the interference between different lossy coupling channels that connect two modes~\cite{Wanjura2020,Huang2021, Huang2023}. Nonreciprocal transmission is realized by combining dissipative couplings with periodic structure, and the nonreciprocity ratio is exponentially enhanced by increasing the chain length. Furthermore, this mechanism enables unidirectional energy transmission without introducing additional insertion loss, as the non-Hermiticity induced by dissipative couplings can be fully transformed into nonreciprocity-type non-Hermiticity without bringing extra gain-loss-type non-Hermiticity.

    
\begin{figure}
    \includegraphics[width=\columnwidth]{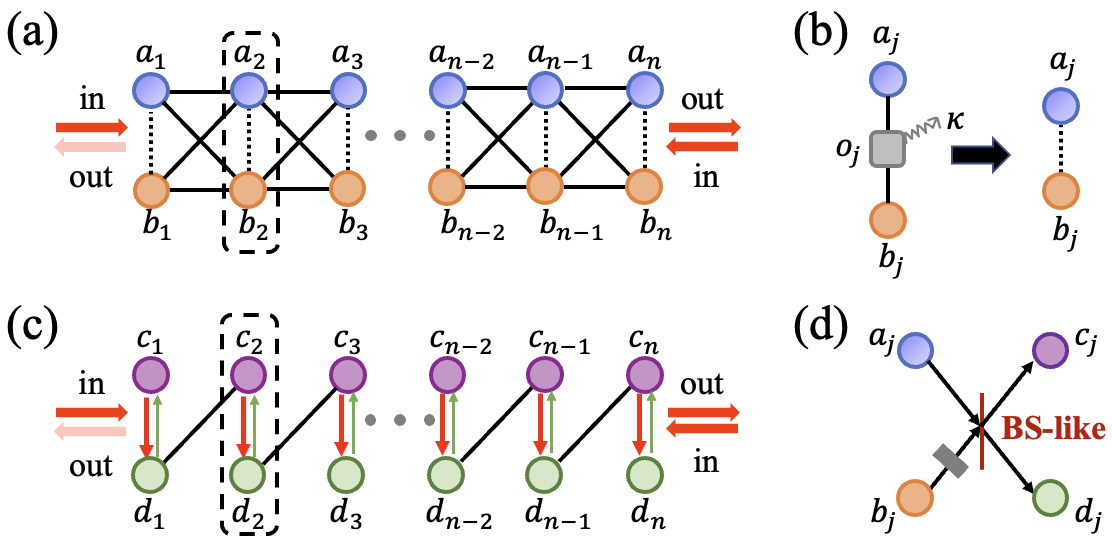}
    \caption{(a) System composed of a chain of resonance modes with dissipative couplings ($iv$, dotted lines) in each unit cell (illutrated by dashed  lines) and intercell coherent couplings (solid lines). The input and output fields are illustrated by the red arrows. (b) One typical example of dissipative coupling implementation by interacting the mode $a_{j}$ and $b_{j}$ with an auxiliary mode $o_{j}$ with large dissipation rate $\kappa$. (c) Non-Hermitian SSH model with intracell unequal left-right couplings ($\delta\pm v$) and intercell coherent  couplings. (d) The equivalence of the two models depicted in (a) and (c) can be understood by interfering the two modes $a_{j}$ and $b_{j}$ using a beam splitter with the phase shift being $-\pi/2$. }
    \label{fig1}
    \end{figure}

    \textit{Model.---}
    As illustrated in Fig.~\ref{fig1}(a), we consider a chain of resonance modes that linearly interact with each other. 
    The couplings can be divided into two types, i.e., the intercell coherent couplings and the dissipative couplings in each unit cell. One typical example of dissipative coupling implementation is depicted in Fig.~\ref{fig1}(b), where mode $a_{j}$ and $b_{j}$ are indirectly coupled via an intermediate lossy mode $o_{j}$. By adiabatically eliminating the lossy mode $o_{j}$, the dissipative coupling term has the form $iv(a_{j}^{\dag}b_{j}+a_{j}b_{j}^{\dag})$, and $v$ can be a real number under the condition that the decay rate of $o_{j}$ is large enough~\cite{Yang2017,SM}. It is non-Hermitian due to the irreversible energy flow in the coupling process, which is different from the coherent couplings that the energy is conserved. 
    The total  Hamiltonian of the system demonstrated in Fig.~\ref{fig1}(a) is 
    \begin{equation}
    \begin{aligned}
    H&=\sum_{j=1}^{n} [\delta(a_{j}^{\dag}a_{j}-b_{j}^{\dag
    }b_{j})+iv(a_{j}^{\dag}b_{j}+a_{j}b_{j}^{\dag})]\\
    &+\sum_{j=1}^{n-1}(w_{1}b_{j}^{\dag}a_{j+1}+w_{2}a_{j}^{\dag}b_{j+1}+H.c.)\\
    &+\sum_{j=1}^{n-1}(u_{1}a_{j}^{\dag}a_{j+1}+u_{2}b_{j}^{\dag}b_{j+1}+H.c.),
    \end{aligned}
    \label{eq1}
    \end{equation}
    where $\delta$ is the difference of the mode resonance frequencies. The intercell coherent couplings contain the coupling between $a_{j}$ and $a_{j+1}$ ($b_{j}$ and $b_{j+1}$) with the rate being $u_{1}$ ($u_{2}$), and the coupling between $b_{j}$ and $a_{j+1}$ ($a_{j}$ and $b_{j+1}$) with the coefficient being $w_{1}$ ($w_{2}$).  
    Since the on-site loss term come from the adiabatic elimination of the intermediate lossy modes $o_{j}$ only shifts the imaginary part of the all eigenfrequencies to the value $iv$ axis,  we have ignored the global on-site loss term in Eq.~\eqref{eq1} for the analysis of the energy spectrum given below. However, we have considered the effect of the global on-site loss term on the dynamical evolution of the modes for calculating the transmission efficiencies (the details can be found in the supplemental material~\cite{SM}).

    We show that under a unitary transformation, our model is equivalent to the non-Hermitian SSH model with unequal left-right couplings. Applying the Fourier transformation $a_{j}=\frac{1}{\sqrt{n}}\sum\nolimits_{k}a_{k}e^{ikj}$, $a_{j}^{\dagger}=\frac{1}{\sqrt{n}}\sum\nolimits_{k}a_{k}^{\dagger}e^{-ikj}$, the system Hamiltonian $H$ [Eq.~\eqref{eq1}] in momentum ($k$) space can be written as $H_{k}=d_{0}I+d_{x}\sigma_{x}+d_{y}\sigma_{y}+d_{z}\sigma_{z}$, where $I$ is the identity matrix and $\sigma_{x,y,z}$ are Pauli matrices. The parameters $d_{0,x,y,z}$ in $H_{k}$ are written as 
    \begin{equation}
        \begin{aligned}
        d_{0}&=\text{Re}(u_{1}+u_{2})\cos k-\text{Im}(u_{1}+u_{2})\sin k,\\
        d_{z}&=\delta+\text{Re}(u_{1}-u_{2})\cos k-\text{Im}(u_{1}-u_{2})\sin k,\\
        d_{x}&=iv+\text{Re}(w_{1}+w_{2})\cos k-\text{Im}(w_{1}+w_{2})\sin k,\\
        d_{y}&=\text{Re}(w_{1}-w_{2})\cos k+\text{Im}(w_{1}-w_{2})\sin k.
        \end{aligned}
        \end{equation}
    Considering the case that the coherent coupling coefficients satisfy $w_{1}=w_{2}=w$ being real and $u_{1}=-u_{2}=iw$, we can simplify the parameters as $d_{0}=0$, $d_{x}=iv+2w\cos k$, $d_{y}=0$, and $d_{z}=\delta-2w\sin k$. Employing the unitary transformation $U_{k}=[1,1;i,-i]/\sqrt{2}$, $H_{k}$ can be transformed as
    \begin{equation}
    H'_{k}=U_{k}^{-1}H_{k}U_{k}=\left(\begin{array}
    [c]{cc}%
     0& \delta+v-2iwe^{-ik}\\
    \delta-v+2iwe^{ik} & 0%
    \end{array}\right).
    \end{equation}
    The off-diagonal terms become non-Hermitian due to the presence  of the dissipative couplings ($|v|\neq0$). To find the model whose momentum-space Hamiltonian is given by $H'_{k}$, we transform $H'_{k}$ back to the real space and get
    \begin{equation}
    \begin{aligned}
    H'&=\sum_{j=1}^{n} [(\delta+v)c_{j}^{\dag}d_{j}+(\delta-v)c_{j}d_{j}^{\dag}]\\
    &+\sum_{j=1}^{n-1}2iw(d_{j}^{\dag}c_{j+1}-c_{j+1}^{\dag}d_{j}),
    \end{aligned}
    \end{equation}
    \label{eq:nssh-H}where the site modes are labeled by $c_{j}$ and $d_{j}$. $H'$ describes a one-dimensional non-Hermitian SSH model with unequal left-right coupling strengths $\delta\pm v$ in each unit cell [Fig.~\ref{fig1}(c)].  
    To establish its connection with our model, we can find that they are equivalent by performing the unitary transformation $H'=U^{-1}HU=U^{\dagger}HU$, where the transformation matrix in real space is the direct sum of that in momentum space, i.e., $U=\oplus_{i=1}^{n} U_{k}$. 
    The equivalence between the two models can be undertood as the multichannel interference provided by the periodic dissipative-coherent coupling structure can lead to unequal left-right coupling between mode $c_{j}=(a_{j}-ib_{j})/\sqrt{2}$ and $d_{j}=(a_{j}+ib_{j})\sqrt{2}$. The resulting asymmetrically coupled modes $c_{j}$ and $d_{j}$ are equivalent to interfer the  modes $a_{j}$ and $b_{j}$ using a beam splitter with the phase factor being $-\pi/2$ [Fig.~\ref{fig1}(d)].
  
    \begin{figure*}
        \includegraphics[width=2\columnwidth]{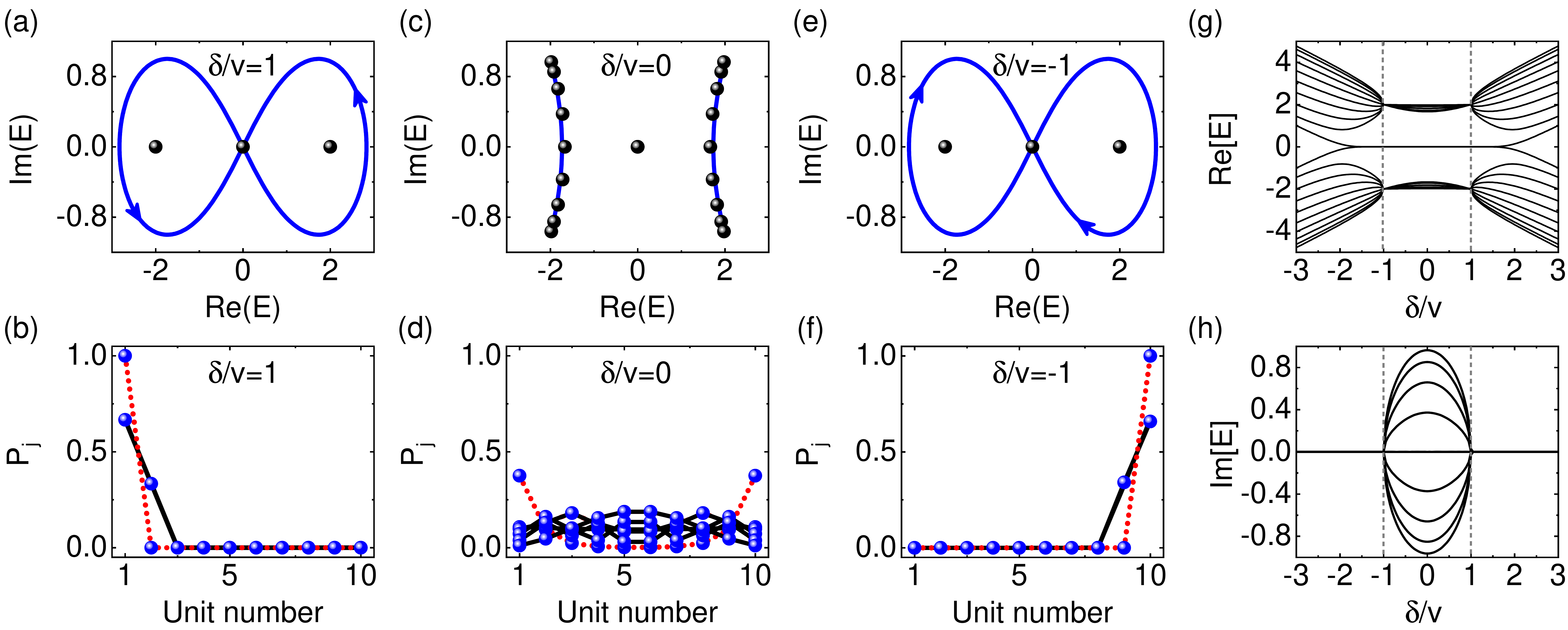}
        \caption{The energy spectrum under PBC (blue lines) and OBC (black dots) are illustrated when $\delta/v=1$ (a), $\delta/v=0$ (c) and $\delta/v=-1$ (e). The open-boundary spectrum is degenerate at $E=\pm2$ and $E=0$ in (a) and (c). The corresponding spatial profile of all eigenmodes under OBC are shown in (b), (d) and (f), respectively. The results obtained from the equivalent model with unequal left-right couplings are depicted by blue dots. The red lines illustrate the zero-energy edge modes. The real (g) and imaginary (h) part of the energy spectrum under OBC as functions of $\delta/v$ are also plotted. Other parameters are fixed as $n=10$, $u_{1}=-u_{2}=iw$, $|w/v|=1$, and $v=-1$.}
        \label{fig2}
     \end{figure*} 

However, different from the non-Hermiticity induced by unequal left-right couplings, our system with dissipative couplings preserves anti-$\mathcal{PT}$ symmetry.
    By defining the parity operation as $\mathcal{P}=\oplus_{i=1}^{n} \sigma_{x}$ with $n$ being the number of the unit cells in the chain, we find that the system Hamiltonian $H$ [Eq.~\eqref{eq1}] preserves anti-$\mathcal{PT}$ symmetry, i.e., $(\mathcal{PT})H(\mathcal{PT})^{-1}=-H$, where $\mathcal{T}$ is the time-reversal operator.
    Different from the global parity operation, $\mathcal{P}$ defined here is a local operation that exchanges the locations of modes in each unitcell, allowing skin effect along the edges. Note that our model is also distinct from the $\mathcal{PT}$ symmetric systems with on-site gain and loss. Anti-$\mathcal{PT}$ symmetry represents a generalization of $\mathcal{PT}$ symmetry that can exist in purely lossy systems. 
    From this point, dissipative couplings provide a distinct origin of non-Hermiticity compared with that induced by unequal left-right couplings and gain-loss systems. 

    \textit{Non-Hermitian skin effect.---}
The energy spectra under periodic boundary condition (PBC, blue lines) and OBC (OBC, black dots) as $\delta/v=1$ are illustrated in Fig.~\ref{fig2}(a). The periodic specrum forms two close loops on the complex plane, which is distinct from open-boundary counterpart that is degenerate at $E=\pm2$ and $E=0$. The real and imaginary part of the open-boundary spectrum as functions of $\delta/v$ are plotted in Fig.~\ref{fig2}(g) and (h), respectively. The eigenvalues are purely real and the bands become degenrate at the exceptional ponts $\delta/v=\pm1$. 
The nonzero winding number of energy under PBC indicates the emergence of NHSE under OBC~\cite{Zhang2020,Okuma2020,Zhang2022a}. For any base point $E_{b}$, the winding number of energy is given by $W=\frac{1}{2\pi i}\int_{-\pi}^{\pi}\partial_{k}\text{ln det}[H(k)-E_{b}]dk$.
This can be seen from Fig.~\ref{fig2}(b), where the sum of the eigenmode population in each unit cell are localized at the boundary. 
The direction of NHSE can also be found from the winding direction. As shown in Fig.~\ref{fig2}(a), the anti-clockwise circling corresponds to $W=1$, resulting in the left localization of the eigenmodes [curves in Fig.~\ref{fig2}(b)].
It agree with the results obtained from the equivalent model with unequal left-right couplings [dots in Fig.~\ref{fig2}(b)], as $\delta/v=1$ corresponds to unidirectional backward coupling. 

The localization direction of the eigenmodes can be reversed by changing the sign of $\delta/v$. As shown in Fig.~\ref{fig2}(e), the spectrum under PBC and OBC for $\delta/v=-1$ are similar to the results when $\delta/v=1$ [Fig.~\ref{fig2}(a)]. However, the winding direction is changed to be clockwise ($W=-1$), indicating that the eigenmodes are localized at the right edge, as depicted in Fig.~\ref{fig2}(f). The results are in accord  with that obtained from the equivlanet model, as $\delta/v=-1$ corresponds to the unidirectional forward coupling.
Figure~\ref{fig2}(c) plots the spectra under PBC and OBC when tuning $\delta/v=0$. Since the spectra coincide and become open arcs, no base point $E_{b}$ can be found to be encircled on the complex plane, indicating that $W= 0$ and NHSE will not exist. It can also be seen from the equivalent model, where $\delta/v=0$ corresponds to equal left-right couplings. Therefore, all the eigenmodes are distrubited throughout the chain as illustrated in Fig.~\ref{fig2}(d).

\begin{figure}
    \includegraphics[width=\columnwidth]{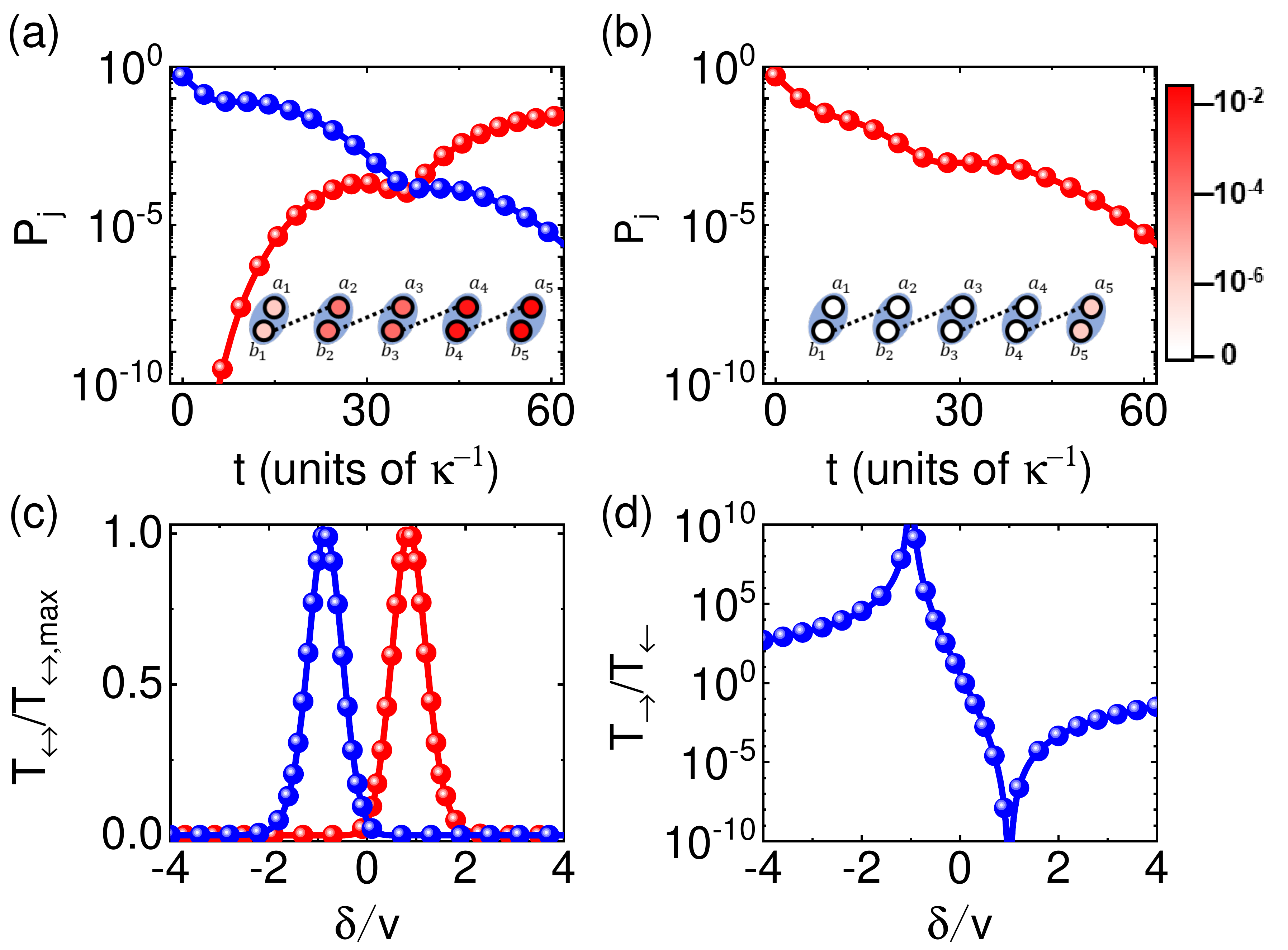}
    \caption{
 Time evolution of the energy occupation for the modes in the leftmost unit (blue) and the rightmost unit (red) in the case of unidirectional forward coupling ($\delta/v=-1$) when the field initially populated in the leftmost unit  (a) and the rightmost unit (b). The energy occupation of $a_{1}$ ($a_{5}$) and $b_{1}$ ($b_{5}$) are plotted by dots and lines, respectively. For backward transmission, the energy occupation of $a_{1}$ and $b_{1}$ is 0 [not shown in (b)] when the initial energy is populated in the rightmost unit. The insets demonstrate the field distribution in the chain at time $t=60\kappa^{-1}$ when the field is transmitted in the forward (a) and backward (b) direction. The field amplitudes of the modes in the unit (illustrated by the blue region) are the same. (c-d) Normalized energy transmission efficiency $T_{\leftrightarrow}/T_{\leftrightarrow,\max}$ and the corresponding nonreciprocity ratio $T_{\rightarrow}/T_{\leftarrow}$ as functions of the ratio $\delta/v$. The analytical and numerical solutions are presented by curves and dots, respectively. The left parameters are fixed as $v/\gamma=-1$, $|w/v|=0.5$, $\gamma/\kappa=0.1$ and $n=5$.}
    \label{fig3}
    \end{figure}

\textit{Nonreciprocity.---}
Nonreciprocity can be realized by combining dissipative couplings with multichannel interference provided by the nonzero synthetic magnetic flux in the periodic structure.
In Fig.~\ref{fig3}(a) and (b), we plot the time evolution of the energy transmission in the case of unidirectional forward coupling ($\delta/v=-1$). 
     When considering the energy is initially occupied in the leftmost unit, i.e., $|a_{1}(t=0)|^{2}=|b_{1}(t=0)|^{2}=1/2$, the energy can transmit in forward direction and the energy occupation of the modes in each unit [inset of Fig.~\ref{fig3}(a)] are the same, i.e., $|a_{i}(t)|^{2}=|b_{i}(t)|^{2}$. However, when the energy is initially occupied in the rightmost unit, i.e., $|a_{n}(t=0)|^{2}=|b_{n}(t=0)|^{2}=1/2$, the energy transmission in backward direction becomes forbidden [Fig.~\ref{fig3}(b)].
     Similarly,  unidirectional backward energy transmission can be obtained by tuning $\delta/v=1$. 
     We can also find that the energy occupation for the modes will decay with time evolving due to the presence of on-site loss.
    
The property of nonreciprocity can also be reflected on the asymmetric scattering matrix of the system~\cite{Jalas2013,Caloz2018,Asadchy2020}. The corresponding steady-state transmission when considering the continuous input field will not be affected by the on-site loss and determined by the parameter matching condition. At the resonance point,  the off-diagonal elements of the scattering matrix $S$ are given as $S_{i,j}=i\sqrt{\gamma_{i}\gamma_{j}}(H+M_{\gamma})^{-1}_{i,j}$,
    where $\gamma_{i}(\gamma_{j})$ is the damping rate associated with the output (input) field.  
    Based on the knowledge that the modes in each unit have the same energy occupation [the insets of Fig.~\ref{fig3}(a) and (b)], we choose the leftmost and rightmost unit to connect the input and output field. Assuming the system dissipation matrix associated with the input/output field as $M_{\gamma}=\text{Diag}[-i\gamma/2, -i\gamma/2, 0, ... -i\gamma/2, -i\gamma/2]$ for simplicity, we can define the energy of the output field as $P_{\text{out}}=|(a_{n}^{\text{out}}+b_{n}^{\text{out}}e^{i\theta})|^{2}/2$, where $\theta$ is the relative phase of the two input (output) fields for forward transmission.  The elements of the scattering matrix  can be solved analytically by applying the relation $H+M_{\gamma}=U_{r}(H_{\text{ns}}+M_{\gamma})U_{r}^{-1}$. We can maximize the forward energy transmission efficiency as $T_{\rightarrow}=P_{\text{out}}/P_{\text{in}}=4|S_{2n,1}|^{2}$ by optimizing the phase factor $\theta$. Similar procedure can be performed to maximize the backward energy transmission efficiency as $T_{\rightarrow}=4|S_{1,2n}|^{2}$. The off-diagonal elements ($S_{2n,1}, S_{1,2n}$) of the scattering matrix are proportional to the unequal left-right coupling coefficients ($\delta\mp v$), which leads to asymmetric forward and backward transmission coefficients (the detailed calculation can be found in the supplemental material~\cite{SM}).
    
 normalized forward and backward energy transmission efficiency ($T_{\leftrightarrow}/T_{\leftrightarrow,\max}$) are illustrated in Fig.~\ref{fig3}(c). The dots and curves represent the numerical and analytical solutions of the scattering matrix $S$, respectively. Unidirectional forward or backward energy transmission can be achieved by tuning $\delta/v$ and the efficiency reaches its maximum when choosing $\delta/v=\mp 1$, i.e., unidirectional coupling.  
    The corresponding nonreciprocity ratio $T_{\rightarrow}/T_{\leftarrow}$ is also shown in Fig.~\ref{fig3}(d). It agrees well with its analytical solution, that is derived as
    \begin{equation}
    \frac{T_{\rightarrow}}{T_{\leftarrow}}=\left(\frac{\delta-v}{\delta+v}\right)^{2n-4} \left(\frac{(v-\gamma/2)^{2}+(\delta-v)^{2}}{(v-\gamma/2)^{2}+(\delta+v)^{2}}\right)^{2}.
    \label{eq:nratio}
    \end{equation}
    Apparently, the nonreciprocity ratio $T_{\rightarrow}/T_{\leftarrow}$ is exponentially enhanced by increasing the unit number $n$ of the chain.  Unidirectional coupling, i.e., $\delta/v=\pm1$ corresponds to tuning the nonreciprocity ratio being 0/infinity. 
    
\begin{figure}
    \includegraphics[width=\columnwidth]{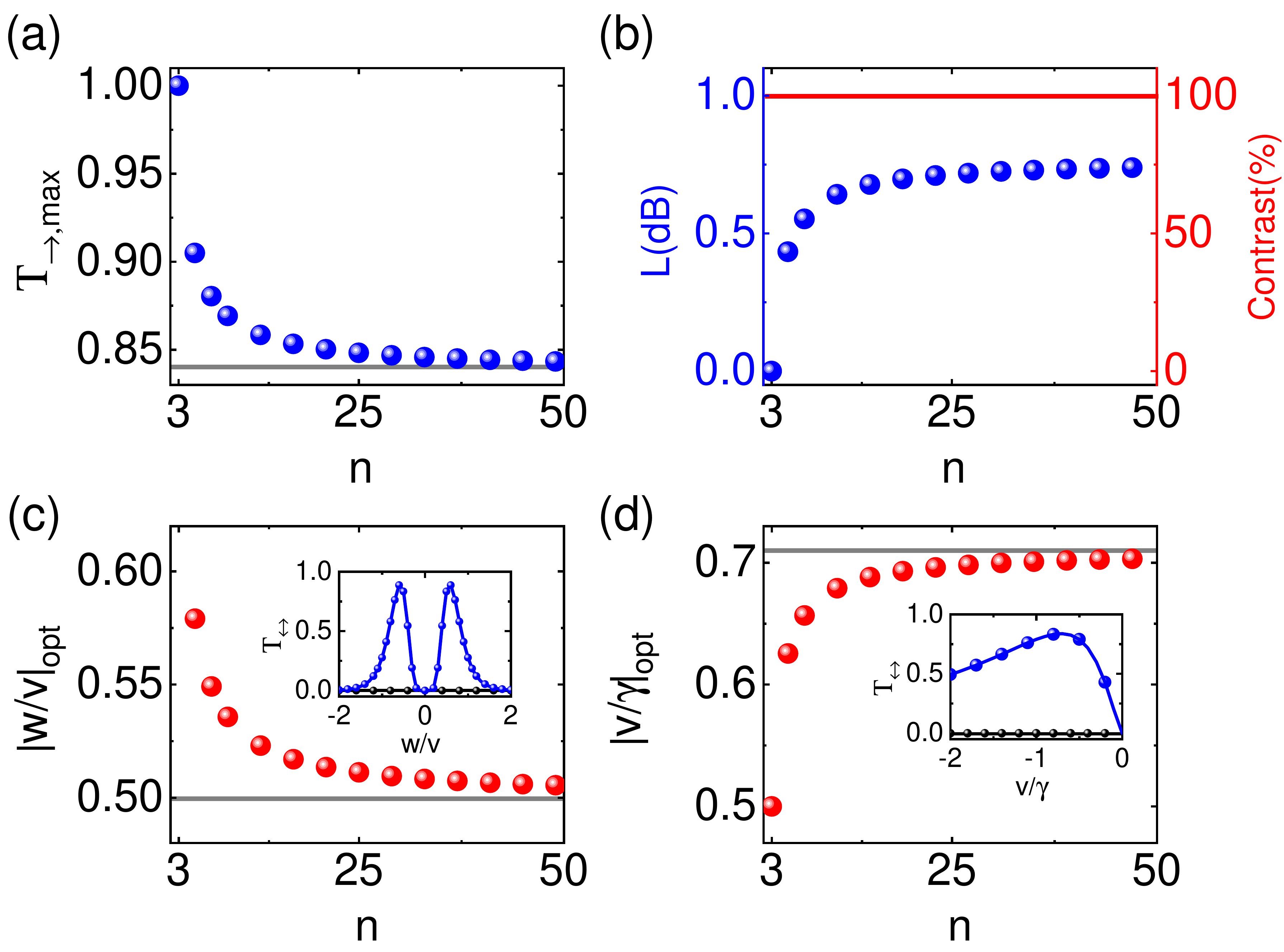}
    \caption{ (a) Unidirectional forward transmission efficiency $T_{\rightarrow}$ as a function of the unit number $n$ for optimized coupling strength ratio $|w/v|$ and coupling-loss ratio $|v/\gamma|$. $T_{\rightarrow,\max}\approx0.84$ with infinite N. (b) The corresponding insertion loss $L$ (dB) and contrast (\%) for unidirectional forward energy transmission. The minimized insertion loss $L\approx 0.76$ dB with infinite $n$. (c-d) Optimal coupling strength ratio $|w/v|$ and coupling-loss ratio $|v/\gamma|$ required to achieve the maximal unidirectional transmission efficiency $T_{\rightarrow,\max}$.  The insets illustrate the transmission efficiencies $T_{\rightarrow}$ (blue) and $T_{\leftarrow}$ (black) as functions of the ratio $w/v$ and $v/\gamma$ when optimizing the left parameters for $n=5$. The analytical and numerical results are presented by curves and dots, respectively.}
    \label{fig4}
    \end{figure}

    As shown in Fig.~\ref{fig4}(a), we maximize the forward transmission efficiency $T_{\rightarrow}$ as a function of the unit number $n$ of the chain in the case of unidirectional forward coupling $\delta/v=-1$. The optimal conditions required for maximizing $T_{\rightarrow}$ are demonstrated in Fig.~\ref{fig4}(c) and (d). 
    Remarkably, we find that with increasing the unit number $n$, $T_{\rightarrow,\max}$ reaches its limit $T_{\rightarrow,\max}\approx0.84$. It can be understood as in the limit of large unit number $n$, complete unidirectional energy transfer between the neighbor units can be realized by optimizing the ratio of the coherent and dissipative coupling strength $|w/v|$ to be $1/2$. Plugging this condition into the expression of $T_{\rightarrow}$, we can find $T_{\rightarrow}$ becomes independent of the unit number $n$ and reach its maximum by further optimizing the coupling-damping rate ratio $|v/\gamma|\approx 0.71$.
    The corresponding insertion loss ($L$) is depicted in Fig.~\ref{fig4}(b), which is below 0.76dB for large unit number $n$. 
We thus conclude that the non-Hermiticity induced by dissipative couplings can be fully transformed into nonreciprocity-type non-Hermiticity without bringing extra gain-loss-type non-Hermiticity. Thus, this mechanism will not introduce extra insertion loss for the unidirectional transmission, enabling the achievement of the nonreciprocal transmission with high effeciency and high contrast simultaneously.

    \textit{Conclusion.---}
    In conclusion, we show that NHSE as well as nonreciprocity can be induced by combining dissipative couplings with periodic structure. By investigating a chain of resonant modes with intracell dissipative couplings and intercell coherent couplings, we find that under a basis change, it can be transformed into a non-Hermitian SSH model with unequal left-right couplings. However, different from the non-Hermiticity induced by unequal left-right couplings and on-site gain and loss, the systems with dissipative couplings preserve local anti-$\mathcal{PT}$ symmetry. 
  When optimizing the unidirectional transmission effeciency, we find that the non-Hermiticity induced by dissipative couplings can be fully transformed into nonreciprocity-type non-Hermiticity without bringing extra gain-loss-type non-Hermiticity. It enables unidirectional energy transmission without introducing additional insertion loss.
    Our work provide a protocol for exploring non-Hermitian topological properties and designing high-performance nonreciprocal devices for one-way optical networks in systems with dissipative couplings.

    \textit{Acknowledgement.---}
    This work is supported by the National Key R\&D Program of China (Grant No. 2023YFA1407600), and the National Natural Science Foundation of China (NSFC) (Grants No. 12275145, No. 92050110, No. 91736106, No. 11674390, No. 91836302, No. 12074027, and No. 12104252).

\bibliographystyle{apsrev4-1}
\bibliography{ref-PerfectNonreciprocity.bib}

\end{document}